\documentclass[aps,showpacs,prl,twocolumn]{revtex4}
\usepackage{graphicx}

\begin{document}

\title{Disorder in a Quantum Critical Superconductor} 
\author{S. Seo$^1$, Xin Lu$^{2,3}$, J.-X. Zhu$^2$, R. R. Urbano$^{2,4}$, N. Curro$^5$, E. D. Bauer$^2$, V. A. Sidorov$^{2,6}$, L. D. Pham$^7$, Tuson Park$^1$, Z. Fisk$^7$, J. D. Thompson$^2$}
\affiliation{$^1$ Department of Physics, Sungkyunkwan University, Suwon 440-746, South Korea \\ $^2$ Los Alamos National Laboratory, Los Alamos, NM 87545, USA \\$^3$ Center for Correlated Matter, Zhejiang University, Hangzhou, 310027, China \\$^4$ Instituto de Fisica “Gleb Wataghin”, Universidade Estadual de Campinas-SP, 13083-859, Brazil \\
$^5$ Department of Physics, University of California, Davis, CA 95616, USA \\
$^6$ Institute for High Pressure Physics, Russian Academy of Sciences, RU-142190 Troitsk, Moscow, Russia \\
$^7$ Department of Physics, University of California, Irvine, CA 92697, USA
}
\date{\today}

\begin{abstract}
In four classes of materials, the layered copper-oxides, organics, iron-pnictides and heavy-fermion compounds, an unconventional superconducting state emerges as a magnetic transition is tuned toward absolute zero temperature, that is, toward a magnetic quantum-critical point (QCP)~\cite{ref1}. In most materials, the QCP is accessed by chemical substitutions or applied pressure.  CeCoIn$_5$ is one of the few materials that are ‘born’ as a quantum-critical superconductor~\cite{ref2, ref3, ref4} and, therefore, offers the opportunity to explore the consequences of chemical disorder. Cadmium-doped crystals of CeCoIn5 are a particularly interesting case where Cd substitution induces long-range magnetic order~\cite{ref5}, as in Zn-doped copper-oxides~\cite{ref6, ref7}. Applied pressure globally supresses the Cd-induced magnetic order and restores bulk superconductivity. Here we show, however, that local magnetic correlations, whose spatial extent decreases with applied pressure, persist at the extrapolated QCP. The residual droplets of impurity-induced magnetic moments prevent the reappearance of conventional signatures of quantum criticality, but induce a heterogeneous electronic state. These discoveries show that spin droplets can be a source of electronic heterogeneity in classes of strongly correlated electron systems and emphasize the need for caution when interpreting the effects of tuning a correlated system by chemical substitution. 
\end{abstract}
\maketitle
Impurities, defects in an otherwise homogeneous host, often are unwanted because their influence can mask intrinsic properties of the host material~\cite{ref8}. Impurities, however, also can be a double-edged sword by providing an avenue to induce interesting new phases of matter and to probe the underlying mechanism of exotic ground states, especially those that emerge from complex interactions in strongly correlated electron materials.  In the high transition temperature ($T_c$) copper-oxide superconductors, the intentional inclusion of small concentrations of non-magnetic impurities, such as Zn, induces magnetism around the impurity site and also enables visualization of the symmetry of the superconducting gap~\cite{ref6, ref7, ref9, ref10}. Nucleation of a charge density wave in regions surrounding impurity sites is another example of defects revealing the underlying competing phase by disorder~\cite{ref11}. With a growing number of classes of materials that show unusual sensitivity to impurities, understanding and controlling the emergent phases from impurities is an important open issue.

Materials that exhibit an extreme sensitivity to impurities often are near a zero-temperature, second order phase transition, where quantum-critical fluctuations of the associated order parameter diverge in space and time. Theory predicts that when disorder is coupled to these critical fluctuations local regions of an exotic phase can nucleate inside the host and change the nature of the phase transition; indeed, even long-range order of the nucleated phase is possible~\cite{ref12, ref13, ref14}. Nuclear magnetic resonance (NMR) and scanning tunnelling microscopy have validated these predictions by showing that the host system locally responds to an impurity by creating an extended droplet of the new phase around an impurity~\cite{ref6, ref15}. Understanding how the local droplet evolves as a function of a control parameter or how the droplet interplays with the host phase, however, is less well known, partly due to lack of very pure compounds that are situated sufficiently close to a quantum-critical point (QCP). The quantum-critical superconductor CeCoIn$_5$ provides an ideal opportunity to probe the consequences of impurities on fluctuations of a quantum-critical state. It forms readily as exceptionally high quality single crystals, and the small (1~meV) characteristic energy of its ground state can be tuned easily with modest applied pressure or magnetic field without introducing additional disorder~\cite{ref16, ref17, ref18}. At ambient pressure, the quantum-critical state of CeCoIn$_5$, characterized by a linear-in-temperature electrical resistivity above $T_c$~\cite{ref3} and  logarithmic divergence of the low-temperature electronic specific heat divided by temperature ($C/T$)~\cite{ref19}, has been interpreted as due to proximity to a field-induced magnetic QCP~\cite{ref16, ref20}. Replacing one atomic percent of In by Cd in CeCo(In$_{1-x}$Cd$_x$)$_5$ reduces $T_c$ from 2.3~K ($x=0$) to $T_c=1.2$~K ($x=0.01$) and induces microscopic coexistence of long–range antiferromagnetic order with $T_N=2.8$~K~\cite{ref5, ref21}. For slightly larger $x$, superconductivity is suppressed completely and only Neel order remains. Introducing these Cd atoms into CeCoIn$_5$ creates defects that produce a response consistent with theoretical expectations~\cite{ref12, ref13, ref14}.
 
Application of pressure accurately reverses the global effect of Cd substitutions, suppressing the long-range magnetic order and inducing a dome of superconductivity~\cite{ref5}. The fact that the global phase response of CeCoIn$_5$ to cadmium doping can be undone by applied pressure suggests that a quantum-critical state should reappear, but the critical point may be hidden by the pressure-induced superconducting state of Cd-doped crystals. This is precisely what is found in the isostructural compound CeRhIn$_5$, which orders antiferromagnetically at atmospheric pressure in the absence of intentionally added impurities. Applying pressure to this correlated electron metal suppresses its long-range antiferromagnetism and induces a dome of bulk superconductivity that hides a QCP which is revealed in an applied magnetic field~\cite{ref17, ref18}. The pressure evolution of $C/T$ in zero-field is displayed in Fig.~1a as a function of temperature for CeCoIn$_5$ doped with 1~\% Cd. Similar results are found for a sample doped with 1.5~\% Cd (see Fig.~4 in Supplementary Information). For comparison, we also plot the pressure dependence of $C/T$ of the pristine reference compound CeRhIn$_5$ in Fig.~1b.  With applied pressure, $T_N$ is suppressed in both CeRhIn$_5$ and Cd-doped CeCoIn$_5$ compounds, and $T_c$ increases. In contrast to CeRhIn$_5$, however, the discontinuity $\Delta C=C-C_N$, where $C_N$ is the normal-state specific heat at $T_c$, normalized by $C_N$, $\Delta C/C_N\approx 1.2$ at $T_c$ for Cd-doped CeCoIn$_5$ at $P=1.21$~GPa. This normalized discontinuity is less than 30~\% of the corresponding value in CeRhIn$_5$ at 2.05~GPa or in pure CeCoIn$_5$ at ambient pressure where these materials have nearly identical $T_c$’s. 
\begin{figure}[tbp]
\centering  \includegraphics[width=8cm,clip]{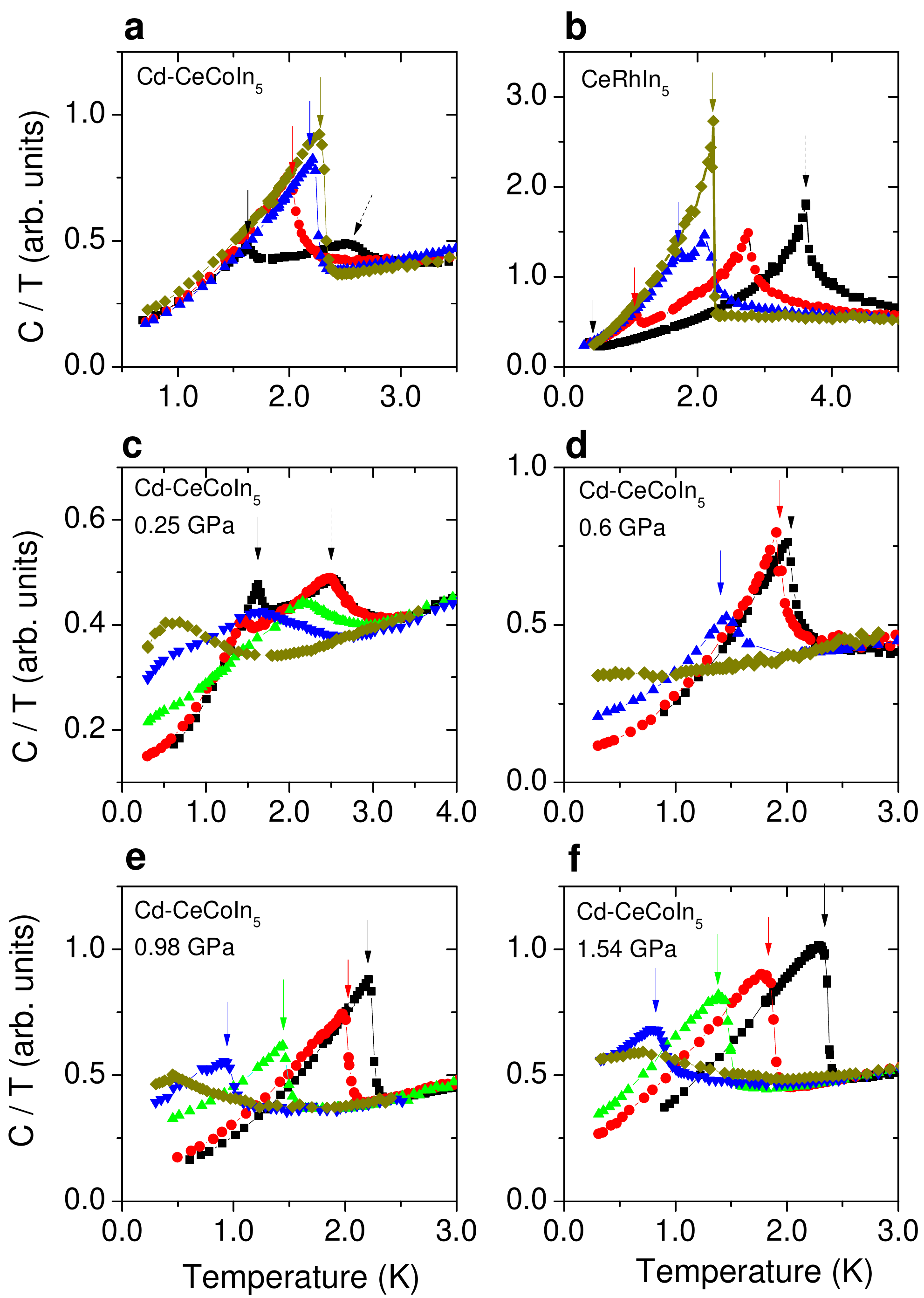}
\caption{Pressure dependence of the specific heat of 1~\% Cd-doped CeCoIn$_5$ and CeRhIn$_5$. \textbf{a}, Specific heat divided by temperature for 1~\% Cd-doped CeCoIn$_5$ in zero applied field and at pressures of 0.25 (black symbols), 0.60 (red), 0.98 (blue), and 1.21~GPa (dark yellow). \textbf{b}, Specific heat divided by temperature for CeRhIn$_5$ at 1.15 (black), 1.51 (red), 1.71 (blue), and 2.05~GPa (dark yellow). In CeRhIn$_5$, the spin entropy of Ce 4f local moments is transferred completely to the superconducting phase when magnetism is suppressed, resulting in a large $C/C_N >4$ at $T_c$~\cite{ref18}. The qualitative difference between CeRhIn$_5$ and Cd-doped CeCoIn$_5$ exists even though the ordered magnetic moment in both is comparable, $\approx 0.7~\mu_B$. \textbf{c-f}, Dependence on temperature of the specific heat divided by temperature of 1~\% Cd-doped CeCoIn$_5$ under magnetic fields and pressures. \textbf{c}, 0.25~GPa at magnetic fields of 0 (black), 1.0 (red), 3.0 (green), 5.0 (blue), and 9.0~Tesla (dark yellow). \textbf{d}, 0.6~GPa at magnetic fields of 0 (black), 1.0 (red), 3.0 (blue), and 5.0~Tesla (dark yellow). \textbf{e}, 0.98~GPa at magnetic fields of 0 (black), 2.0 (red), 4.0 (green), 5.0 (blue), and 5.5~Tesla (dark yellow). \textbf{f}, 1.54~GPa at magnetic fields of 0 (black), 3.0 (red), 4.0 (green), 5.0 (blue), and 6.0~Tesla (dark yellow). Solid and dashed arrows indicate superconducting and antiferromagnetic transition temperatures, respectively. Values of $C/T$ for different pressures were normalized against each other with an assumption that the entropy recovered at 10~K is same for all pressures within each compound. This assumption is proven for Cd-doped CeCoIn$_5$ as a function of Cd content at atmospheric pressure~\cite{ref5}.}
\label{figure1}
\end{figure}

The effect of a magnetic field on the specific heat of the 1~\% Cd-doped CeCoIn$_5$ is summarized in Fig.~1c-f for pressures up to 1.54~GPa. For magnetic fields above the critical field $B_{c2}$, where superconductivity is completely destroyed, $C/T$ at low temperatures decreases with decreasing temperature typical of a non-critical metal.  At 0.98~GPa and 6.2~T, which is just above the upper critical field $B_{c2}$ at this pressure, there is a slight upturn in $C/T$ at the lowest temperatures. This weak upturn, however, is distinct from undoped CeCoIn$_5$, where $C/T$ logarithmically diverges with decreasing temperature near $B_{c2}$~\cite{ref16} and from CeRhIn$_5$ at its quantum-critical pressure, where $C/T$ also diverges once superconductivity is suppressed by a magnetic field~\cite{ref18}. The prominent lack of a divergence in $C/T$ of the Cd-doped compound indicates that quantum-critical behaviour is avoided. 

Supporting this lack of quantum criticality from specific heat measurements, the electrical resistivity $\rho$ of 1~\% Cd-doped CeCoIn$_5$ also does not show an anomalous temperature dependence characteristic of proximity to a QCP. Isothermal cuts of the low-temperature electrical resistivity as a function of pressure (Fig.~2a) show that $\rho(P)$ decreases monotonically for $T \geq T_c$ across the critical pressure $P_{c1}$ ($\approx$ 0.5~GPa), where $T_N$ becomes equal to $T_c$, and $P_{c2}$ ($\approx$1.0~GPa), the extrapolated  critical point where $T_N$ extrapolates to 0~K inside the superconducting dome.  Further, Fig.~2c shows a linear$-T$ resistivity over a temperature range between 12 and 35~K at 1.89~GPa, but it becomes sub-linear in $T$ at lower temperatures, in contrast to expectations of quantum criticality and the $T-$linear $\rho(T)$ in pristine CeCoIn$_5$ that extends to $T_c$~\cite{ref3}. This sub-linear temperature dependence of the resistivity demonstrates that the spectrum of spin excitations has changed with doping such that the scattering rate below $T^*$ is reduced relative to a continuation of the $T-$linear resistivity to lower temperatures, a reduction in scattering arising from possible short range magnetic correlations. 
\begin{figure}[tbp]
\centering  \includegraphics[width=8cm,clip]{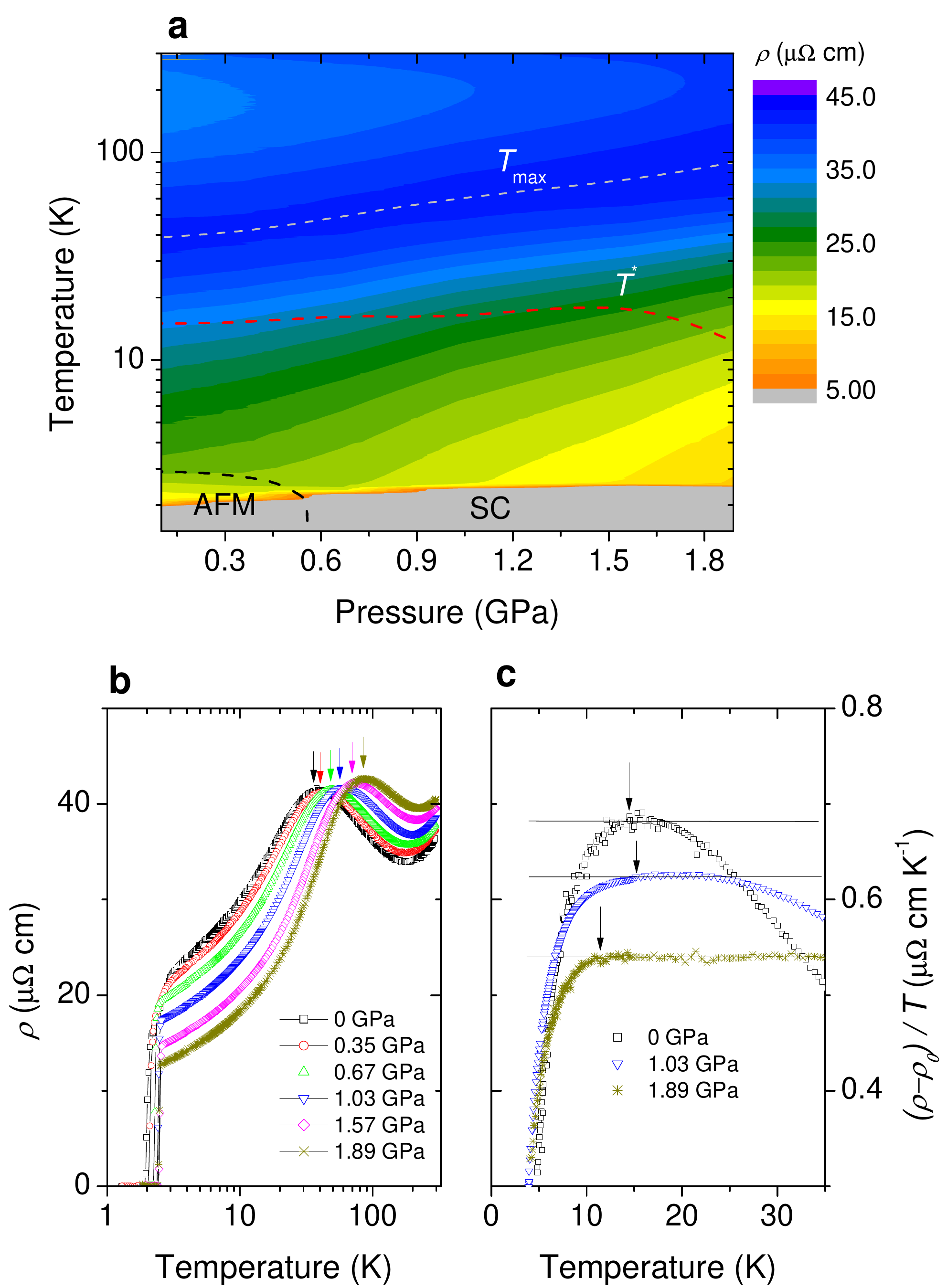}
\caption{Electrical resistivity of 1~\% Cd-doped CeCoIn$_5$ under pressure. \textbf{a}, Contour map of the electrical resistivity ($\rho_{ab}$) within the Ce-In plane is plotted in the pressure-temperature plane. $T_{max}$ is a temperature where the resistivity reaches a maximum, while $T^*$ is the temperature below which the resistivity deviates from a $T-$linear dependence. An antiferromagnetic state (AFM) coexists with superconductivity (SC) for $P < 0.5$~GPa. \textbf{b}, In-plane resistivity $\rho_{ab}$ is shown as a function of temperature on a semi-logarithmic scale for representative pressures: 0 (black), 0.35 (red), 0.67 (green), 1.03 (blue), 1.57 (magenta), and 1.89~GPa (dark yellow). Arrows mark the evolution of $T_{max}$ with pressure. The temperature $T_{max}$ at which $\rho(T)$ is a maximum occurs at 34~K at ambient pressure and increases linearly with increasing pressure, which is typical of strongly correlated Ce-based materials, such as pure CeCoIn$_5$, and reflects a pressure-induced increase of the hybridization between  ligand electrons and the periodic array of Ce 4f-electrons. \textbf{c}, Resistivity, with the residual $T=0$ value subtracted, divided by temperature, ($\rho_{ab} -\rho_{ab}(0T)) /T$, is plotted against temperature for representative pressures 0 (black), 1.03 (blue), and 1.89~GPa (dark yellow). $T^*$ is marked by arrows.}
\label{figure2}
\end{figure}

The large jump in specific heat at $T_c$ in pure CeCoIn$_5$ and in CeRhIn$_5$ at its pressure-tuned QCP is a consequence of the huge quantum-disordered entropy of magnetic fluctuations that is recovered in the zero-temperature limit when superconductivity is suppressed by a magnetic field. In contrast, the specific heat jump in Cd-doped CeCoIn$_5$ is smaller, even when these crystals are tuned to their putative critical point. This small jump, together with the absence of a signature for a diverging specific heat in a magnetic field and $T-$linear $\rho(T)$ to $T_c$, indicates that the spectrum of magnetic fluctuations has changed in response to the presence of Cd impurities. When very dilute concentrations of non-magnetic impurities are added to CeCoIn$_5$, a resonance in the electronic density of state develops from unitary scattering of the highly correlated band of conduction electrons by the local defect~\cite{ref22}. In this process, the non-magnetic defect acquires magnetic character of a spin $S=1/2$ Kondo impurity. At higher concentrations, local droplets of static magnetic order nucleate with a typical extent of a few lattice parameters~\cite{ref21}, and when these droplets overlap, long-range antiferromagnetic order develops. Although magnetic fluctuations may be suppressed locally through the nucleation of spin droplets around the Cd impurities, they should not be affected significantly at distances removed from these impurities where there is no static magnetism. Apparently, sufficient spectral weight remains in the unaffected fluctuations to produce a superconducting transition only slightly lower than that of pure CeCoIn$_5$.

Figure 3a shows $^{115}$In nuclear quadrupolar resonance (NQR) spectra of the In(1) site for CeCoIn$_5$ and 1~\%~Cd-doped CeCoIn$_5$. Three peaks A, B, and C are observed in Cd-doped crystals~\cite{ref21}, where peak A probes the bulk of the unit cells as in pristine CeCoIn$_5$. The existence of B and C peaks underscores the presence of a minority of In sites in an electronic environment not present in undoped CeCoIn$_5$. Cd impurities significantly broaden the In(1) NQR spectra due to the distribution of local electric field gradients. Figure~3b shows the temperature dependence of the spin-lattice relaxation rate $1/T_1$ measured at peak A of 1~\% Cd-doped CeCoIn$_5$ under pressures to 1.51~GPa. At ambient pressure, $1/T_1$ is identical to the pure compound and follows a $T^{1/4}$ temperature dependence in the disordered state, which is expected for a system in close proximity to an antiferromagnetic QCP~\cite{ref21, ref23}. The observation that $1/T_1$ is a strong function of pressure yet independent of Cd doping provides microscopic evidence that pressure is not the reverse of doping. Taken with the specific heat data, these data imply that the electronic response to Cd doping is strongly inhomogeneous. As pressure tunes the system away from long-range magnetic order, the spin-spin correlation length, $\xi$, and therefore the size of the droplets, should decrease. In the temperature range around $T_N$ $(P=0)$, this is reflected in a rapid decrease of the relaxation rate $1/T_1 \propto T\xi^\alpha$~\cite{ref23, ref24, ref25, ref26, ref27} that is plotted in Fig.~3b. These NQR results provide microscopic evidence for the existence of spin droplets even when global long-range antiferromagnetic order is suppressed completely for pressures above $P_{c1}$. 
\begin{figure}[tbp]
\centering  \includegraphics[width=8cm,clip]{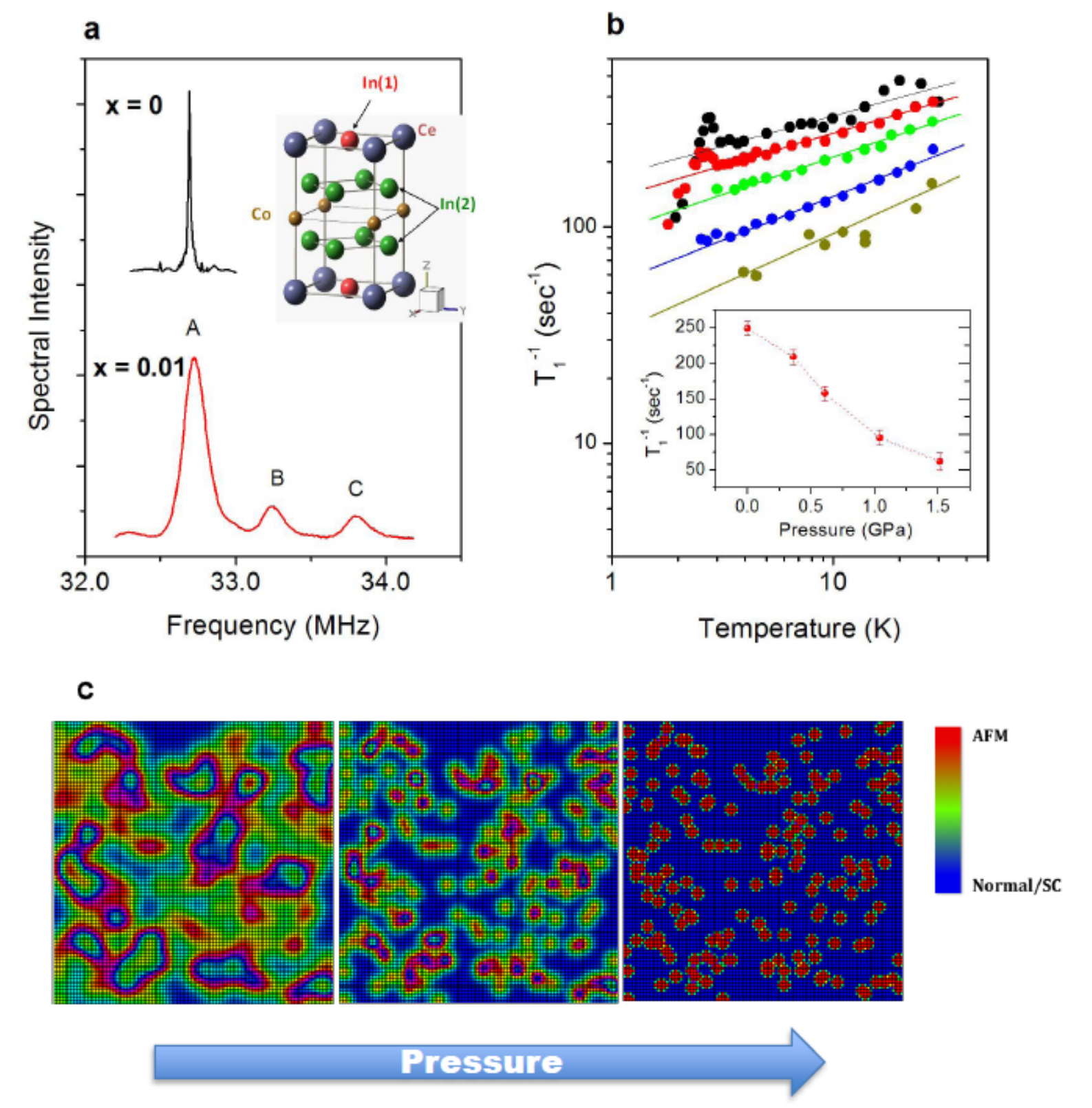}
\caption{(color online) Spin lattice relaxation rate of 1~\% Cd-doped CeCoIn$_5$ under pressure. \textbf{a}, $^{115}$In nuclear quadrupole resonance (NQR) spectra of the In(1) site at ambient pressure. $^{115}$In NQR spectra of CeCoIn$_5$ and 1~\% Cd doped CeCoIn$_5$ are shown in the upper and lower parts of the panel. Indium sites in the unit cell of CeCoIn$_5$ are indicated in the inset. In the Cd-doped crystals, additional peaks B and C appear at higher frequencies, while the original peak A is broadened and moved slightly to higher frequency. \textbf{b}, Dependence on temperature of the spin-lattice relaxation rate $1/T_1$ of CeCo(In$_{0.99}$Cd$_{0.01}$)$_5$. Solid lines are guides to eyes for pressures of 1~bar (black), 0.36 (red), 0.61 (green), 1.04 (blue), and 1.51~GPa (dark yellow). The overall decrease in $1/T_1$, even at high temperatures, is found in pristine CeCoIn$_5$~\cite{ref26}. The inset shows $1/T_1$ at 4~K as a function of pressure. Error bars describe the uncertainty in $1/T_1$. \textbf{c}, Schematic illustration of the dependence on pressure of the size of magnetic droplets. The Cd atoms that replace In in CeCoIn$_5$ nucleate antiferromagnetic droplets surrounding the Cd site. The average distance between the droplets, which is approximately five lattice spacings at 0.75~\% Cd doping, is independent of pressure, but their spatial extent shrinks with pressure.  For a pressure above $P_{c1}$ (the middle panel), the magnetic correlation length becomes shorter than inter-droplet spacing, leading to suppression of the long-ranged antiferromagnetic order.}
\label{figure3}
\end{figure}

The nucleation and pressure evolution of spin droplets can be understood within the framework of a theoretical model that considers the competition between antiferromagnetism and d-wave superconductivity~\cite{ref13} (details of the model are discussed in SI). In the absence of impurities, this model gives negligible spectral weight in the magnetic channel and homogeneous d-wave superconductivity near a QCP. When a unitary scattering impurity is introduced, however, superconductivity is suppressed strongly around the defect but recovers over a distance of a coherence length away from the impurity. Concurrently, magnetic order is depressed at the impurity but is a maximum at the nearest Ce neighbour and decays over many lattice constants. As shown in Fig.~7a and 7b in the Supplementary Information, the superconducting order is robust against a change in effective bandwidth (or the external pressure), but the magnetic order is very sensitive. A slight increase in bandwidth rapidly damps the long-range oscillatory magnetization and the amplitude of the near-neighbour magnetic order parameter. For a dilute concentration of impurity centres or at sufficiently wider bandwidths (higher pressures), there only will be magnetic droplets, but for a narrower bandwidth or higher concentration of scattering centres, magnetic correlations between sites induce global magnetic order. Predictions of this model are illustrated schematically in Fig.~3c. 

The lack of signatures for quantum-critical behaviour in Cd-doped CeCoIn$_5$ manifests a novel mechanism for the coexistence of magnetism and superconductivity and provides a new perspective for interpreting the response to disorder in other strongly correlated superconductors near a zero-temperature magnetic instability. As this work shows, tuning a system with or by disorder to a presumed magnetic QCP does not necessitate a quantum-critical response and associated spectrum of quantum fluctuations.  Though not all impurities may be unitary scatterers, the likelihood is high that they will be strong scatterers if the bandwidth of the host material is sufficiently narrow, as it is in many strongly correlated heavy-fermion compounds, such as CeCoIn$_5$. The freezing of magnetic quantum fluctuations around impurity sites has broader consequences for the nature of electronic heterogeneity that is common to classes of strongly correlated electron systems.

\subsection{Methods}
Single crystals of CeCo(In$_{1-x}$Cd$_x$)$_5$ were synthesized by a standard In-flux technique and their basic physical properties were reported previously~\cite{ref5}. Electrical resistivity, specific heat, and nuclear quadrupolar resonance (NQR) measurements were performed under pressure for samples with Cd concentrations $x=0.01$ and 0.015, where the concentration $x$ is determined by microprobe measurements. A quasi-hydrostatic pressure environment was achieved in a Be-Cu/NiCrAl hybrid clamp-type cell with silicone oil as the transmitting medium and pressure in the cell was determined at low temperatures by inductive measurements of a change in the superconducting transition temperature of Sn that was placed inside the cell~\cite{ref28}. Specific heat measurements were performed by an ac calorimetric technique, where the voltage in Au/Fe(0.07~\%) thermocouple wire, attached to one facet of the crystal, was monitored and converted to a temperature change that is inversely proportional to the specific heat. Details of the ac calorimetric technique can be found elsewhere~\cite{ref29}. A standard four-probe technique was used to measure the electrical resistivity via an LR700 resistance bridge. The spin-lattice relaxation time $1/T_1$ at the A site, which is associated with the bulk high symmetry In(1) sites, was obtained at zero field by measuring nuclear quadrupolar resonance (NQR) of a single crystalline sample.\\

\textbf{Acknowledgements} We thank F. Ronning, M. Vojta and J. Shim for helpful discussions. Work at Los Alamos was performed under the auspices of the U.S. Department of Energy, Office of Science, Division of Materials Science and Engineering and supported in part by the Los Alamos LDRD program. Work at SKKU is supported by NRF grant funded by the Korean Ministry of Education, Science \& Technology (MEST) (No. 2012R1A3A2048816 \& 220-2011-1-C00014).  RRU acknowledges FAPESP (No. 2012/05903-6). VAS acknowledges a support by RFBR Grant 12-02-00376.\\

\textbf{Author Contribution} SS and XL performed the measurements and contributed equally to this work. Correspondence and requests for materials should be addressed to TP (tp8701@skku.edu) or JDT (jdt@lanl.gov).

\section{Supplementary Information}
\textbf{In this supplement, we describe additional data and analyses that support results in the main text. We also discuss model calculations which show that the effective size of spin droplets shrinks with increasing band width.}

Figure 4a shows the specific heat divided by temperature $C/T$ of 1.5~\% Cd-doped CeCoIn$_5$ as a function of temperature for several pressures. The peak in $C/T$ at $T_N$, marked by dashed arrows, is more rounded than in CeRhIn$_5$, but is similar to 1~\% Cd-doped crystals. Such broadening is unexpected from the relatively little disorder introduced by such small concentrations of impurities. With applied pressure, $T_N$ is suppressed and $T_c$ increases, resembling the pressure response of CeRhIn$_5$. Unlike CeRhIn$_5$, however, the peak in $C/T$ at $T_N$ broadens further with applied pressure and the normalized discontinuity $\Delta C/C_N$ at $T_c$ is small, i.e., approximately 25~\% of the undoped compound, where $\DeltaC=C-C_N$ and $C_N$ is the normal-state specific heat at $T_c$ (see Fig.~4b). In CeRhIn$_5$, the spin entropy of Ce 4f local moments is transferred completely to the superconducting phase when magnetism is suppressed, resulting in a large $\Delta C/C_N >4$ at $T_c$ (see Fig.~1b)~[S1]. The small jump in the specific heat of 1.5~\% Cd-doped CeCoIn$_5$ indicates that Cd impurities are consuming a significant fraction of entropy from magnetic degrees of freedom out of which superconductivity develops. 
\begin{figure}[tbp]
\centering  \includegraphics[width=7.5cm,clip]{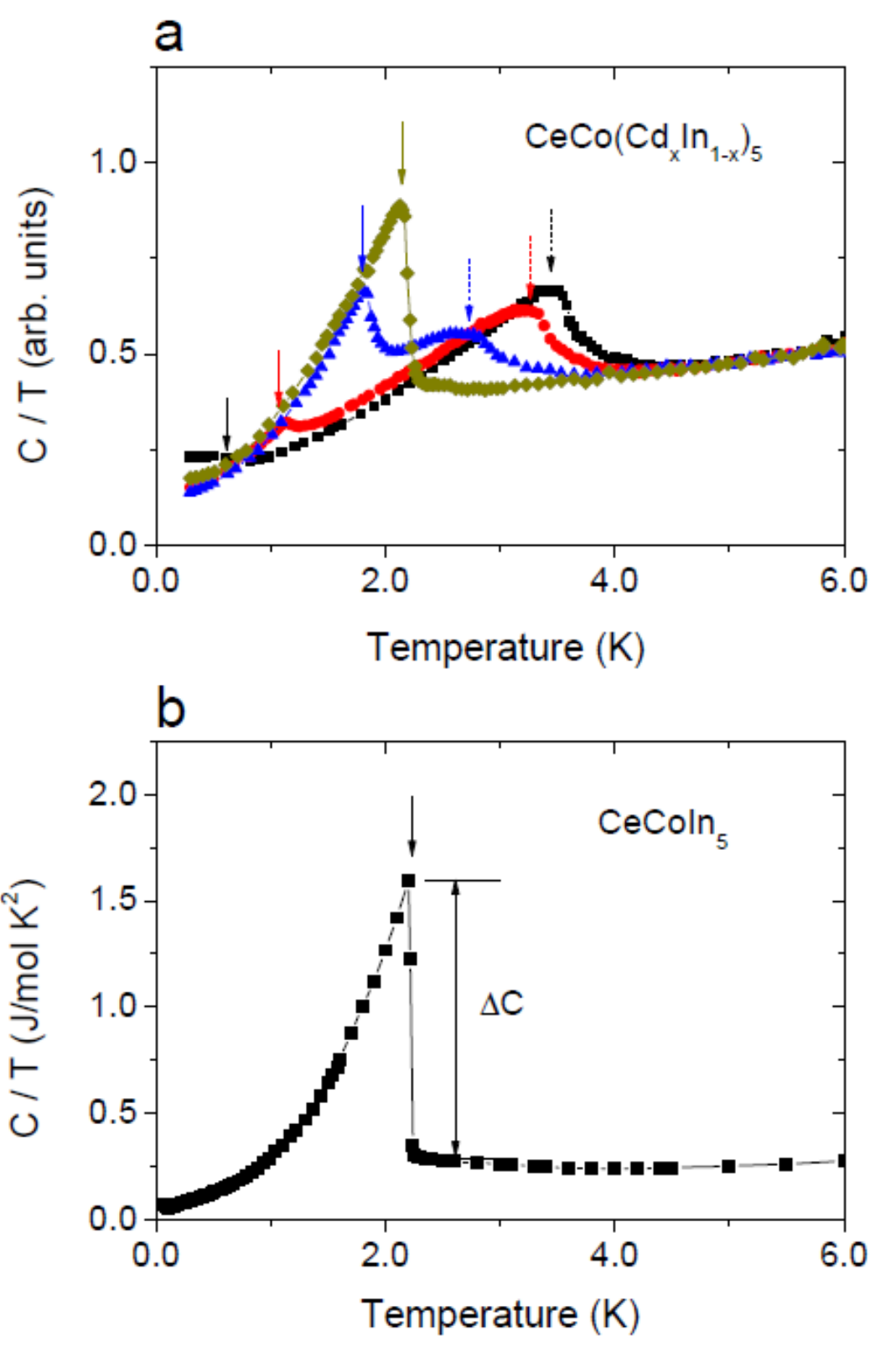}
\caption{ \textbf{a}, Temperature dependence of zero-field specific heat divided by temperature of 1.5~\% Cd-doped CeCoIn$_5$ at pressures of 0.25 (black symbols), 0.60 (red), 0.98 (blue), and 1.21~GPa (dark yellow). Solid and dashed arrows mark superconducting and antiferromagnetic transition temperatures, respectively. Values of $C/T$ for different pressures were normalized against each other with an assumption that the entropy recovered at 10~K is same for all pressures. \textbf{b}, Specific heat divided by temperature $C/T$ of CeCoIn$_5$. The arrow indicates the superconducting phase transition at 2.3~K. $\Delta C$ is the specific heat discontinuity at $T_c$.}
\label{figure 4}
\end{figure}

Figure 5a shows the electrical resistivity of 1.0~\% Cd-doped CeCoIn$_5$ near the superconducting transition temperature. At ambient pressure, $\rho(T)$ has an inflection at 2.9~K where spin disorder scattering is reduced below the antiferromagnetic transition temperature $T_N$. Superconductivity is followed at lower temperature and coexists with the magnetically ordered phase on a microscopic scale below $T_c$ [S2]. Unlike the normal state at room temperature, however, the low-temperature resistivity in the disordered state decreases with pressure, possibly reflecting a decrease in scattering by antiferromagnetic fluctuations. The evolution of ground states with pressure is plotted in Fig.~5b, where $T_N$ is determined from an inflection point and $T_c$ is the zero-resistance temperature. $T_N$ gradually decreases with pressure and is abruptly suppressed above the critical pressure $P_{c1}$ ($\approx$0.5~GPa) where $T_N$ becomes equal to $T_c$. A projected quantum critical point $P_{c2}$ ($\approx$1.0~GPa) was estimated from an extrapolation of $T_N(P)$ to zero temperature. As shown in the main text, however, the lack of a signature for a quantum critical point (QCP) is consistent with nucleation of antiferromagnetic droplets whose effective size decreases as the system moves away from a QCP.
\begin{figure}[tbp]
\centering  \includegraphics[width=7.5cm,clip]{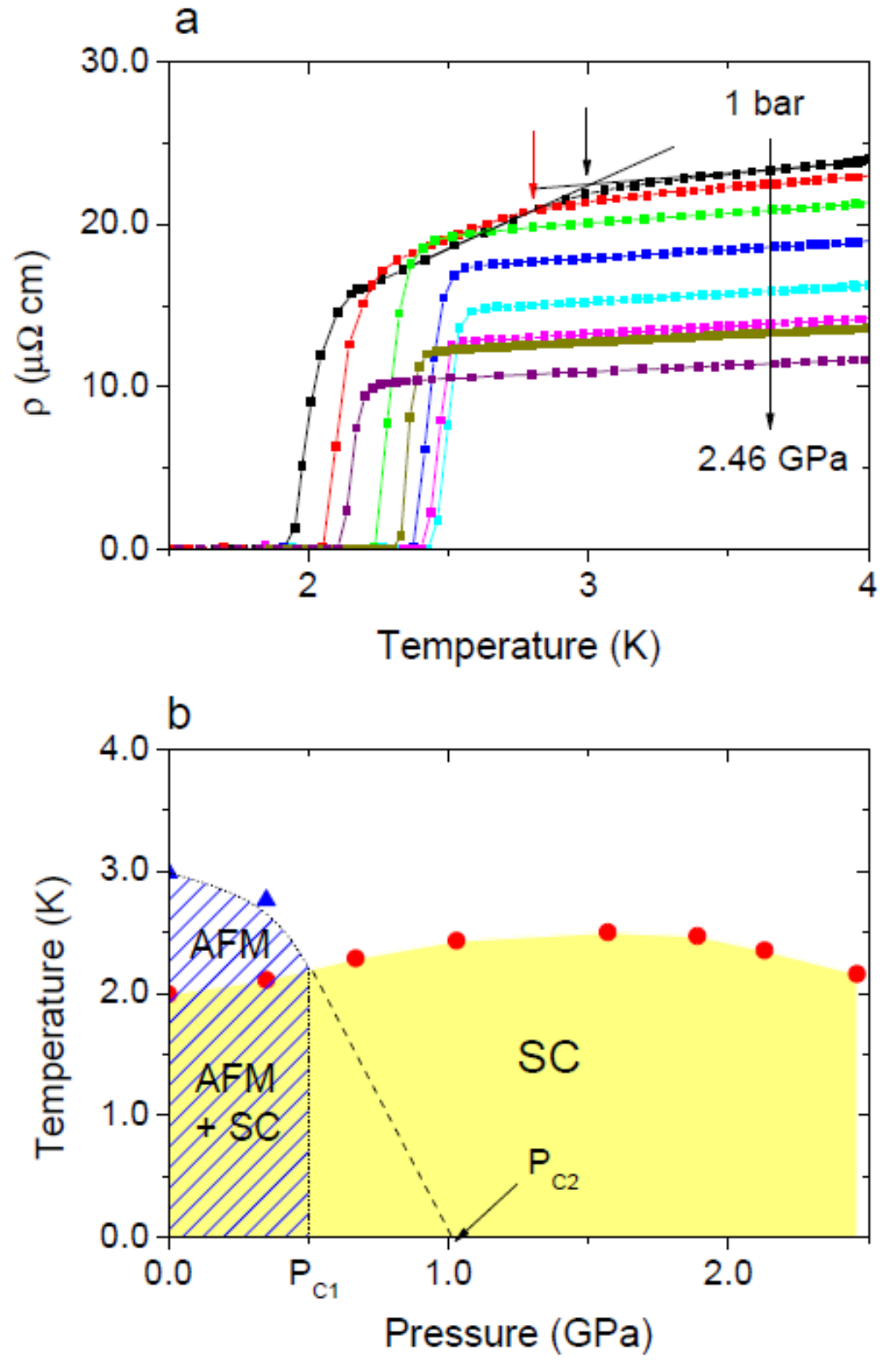}
\caption{ \textbf{a}, Electrical resistivity of 1~\% Cd-doped CeCoIn$_5$ under pressure. Electrical resistivity is plotted against temperature near the superconducting phase transition $T_c$ for pressures of 1~bar (black), 0.35 (red), 0.67 (green), 1.03 (blue), 1.57 (cyan), 1.89 (magenta), 2.13 (dark yellow), and 2.46~GPa (purple). The short arrows indicate the antiferromagnetic transition temperature $T_N$ where there is an inflection in $\rho(T)$. \textbf{b}, Temperature-pressure phase diagram of 1.0~\% Cd-doped CeCoIn$_5$. SC and AFM stand for superconducting and antiferromagnetic states, respectively. $P_{c1}$ ($\approx 0.5$~GPa) is a critical pressure above which AFM is completely suppressed, and $P_{c2}$ ($\approx 1.0$ GPa) is a projected quantum critical point.}
\label{figure 5}
\end{figure}

Figure 6a shows the average distance between Cd impurities and the critical pressure $P_{c1}$ as a function of Cd concentration (\%) on the left and right ordinates, respectively. The average impurity distance was estimated from the inverse cube root of the Cd concentration $(x)^{-1/3}$ in CeCo(In$_{1-x}$Cd$_x$)$_5$ and is plotted in units of the in-plane lattice constant. The critical pressure $P_{c1}$ was determined from resistivity measurements. Error bars on $P_{c1}$ are due to steps between adjacent pressure data sets. From data in Fig.~6a, we estimate the droplet size from the assumption that its size is equal to the average-impurity distance at the critical pressure $P_{c1}$. This assumption is based on an argument from NMR measurements that long-range magnetic order develops when the droplets begin to overlap [S2], and thus at $P_{c1}$ where long-range order disappears, the droplets’ spatial extension just becomes equal to the inter-droplet distance. The pressure evolution of the effective size of spin droplets is plotted in units of lattice constants in Fig.~6b. The droplet size significantly decreases with pressure from 5.22 at ambient pressure to 3.68 lattice constants at 1.3~GPa, while the in-plane Ginzburg-Landau superconducting coherence length increases with pressure [S3]. The contrasting dependence on the pressure of the two length scales suggests that the droplet size is primarily determined by the distance from a QCP.
\begin{figure}[tbp]
\centering  \includegraphics[width=7.5cm,clip]{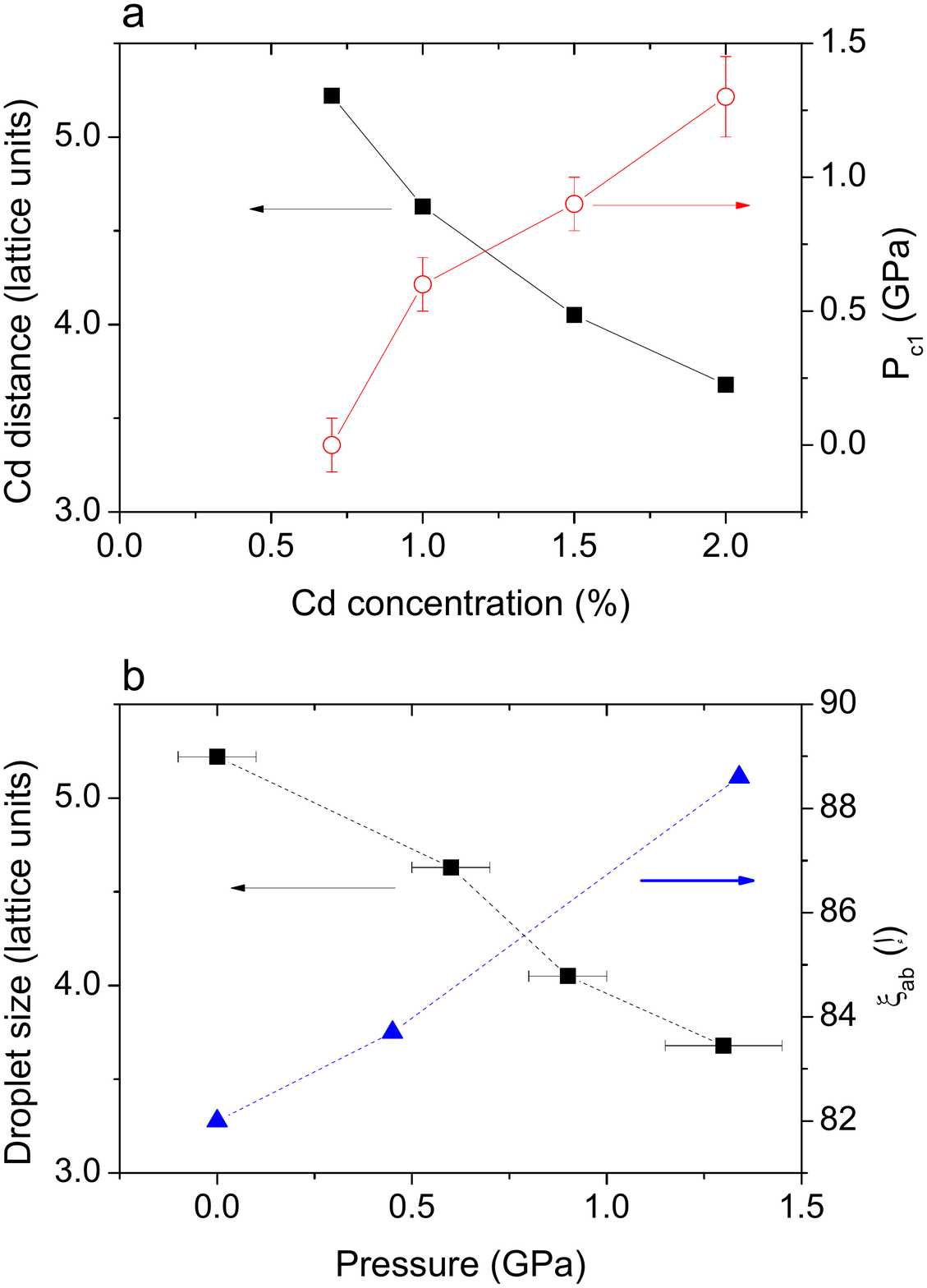}
\caption{\textbf{a}, Average distance between Cd atoms and the critical pressure $P_{c1}$ as a function of Cd concentration (\%) on the left and right ordinates, respectively. Error bars on the critical pressure arise from finite steps between adjacent pressure data sets. \textbf{b}, Pressure dependence of the effective droplet size in units of lattice parameters and the superconducting Ginzburg-Landau coherence length on the left and right ordinates, respectively.}
\label{figure 6}
\end{figure}

Theoretically, we considered a microscopic model with competing interactions. In this model, an on-site repulsion is solely responsible for the antiferromagnetism while a nearest-neighbor attraction generates the d-wave superconductivity [S4]. We note that these interactions are effective, which should be understood to exist on top of a renormalization band formed from a competition in a much more fundamental level in heavy fermion systems. We introduced a single-site potential to describe the nonmagnetic impurity scattering effects. Due to the strong band renormalization, the impurity scattering effects are amplified in heavy fermion systems. The effect of pressure is to enhance the hybridization strength between the local f-electron and extended conduction bands, and therefore increases the bandwidth in our effective band renormalization model. As such we describe the renormalization bandwidth in the form of $\alpha D_0$, where $D_0$ is the one at ambient pressure and the scaling factor $\alpha$ increases with applied pressure. For computational efficiency, numerical simulations were performed on a two-dimensional square lattice. This treatment is a reasonable approximation in view of the fact that the CeCoIn$_5$ compounds have a layered structure, in which the dominant properties occur in the basal planes containing Ce atoms. In the tight-binding model, we used the parameter values at ambient pressure: the nearest and next-nearest neighbor hopping integrals $t=1$ and $t’=-0.3$, the on-site Coulomb repulsion $U=2.78$, the d-wave pairing interaction $V_d=1.4$, the electron filling factor $n\approx$0.90, and the strength of the impurity scattering potential $\epsilon_i=100$.

\begin{figure}[tbp]
\centering  \includegraphics[width=7.5cm,clip]{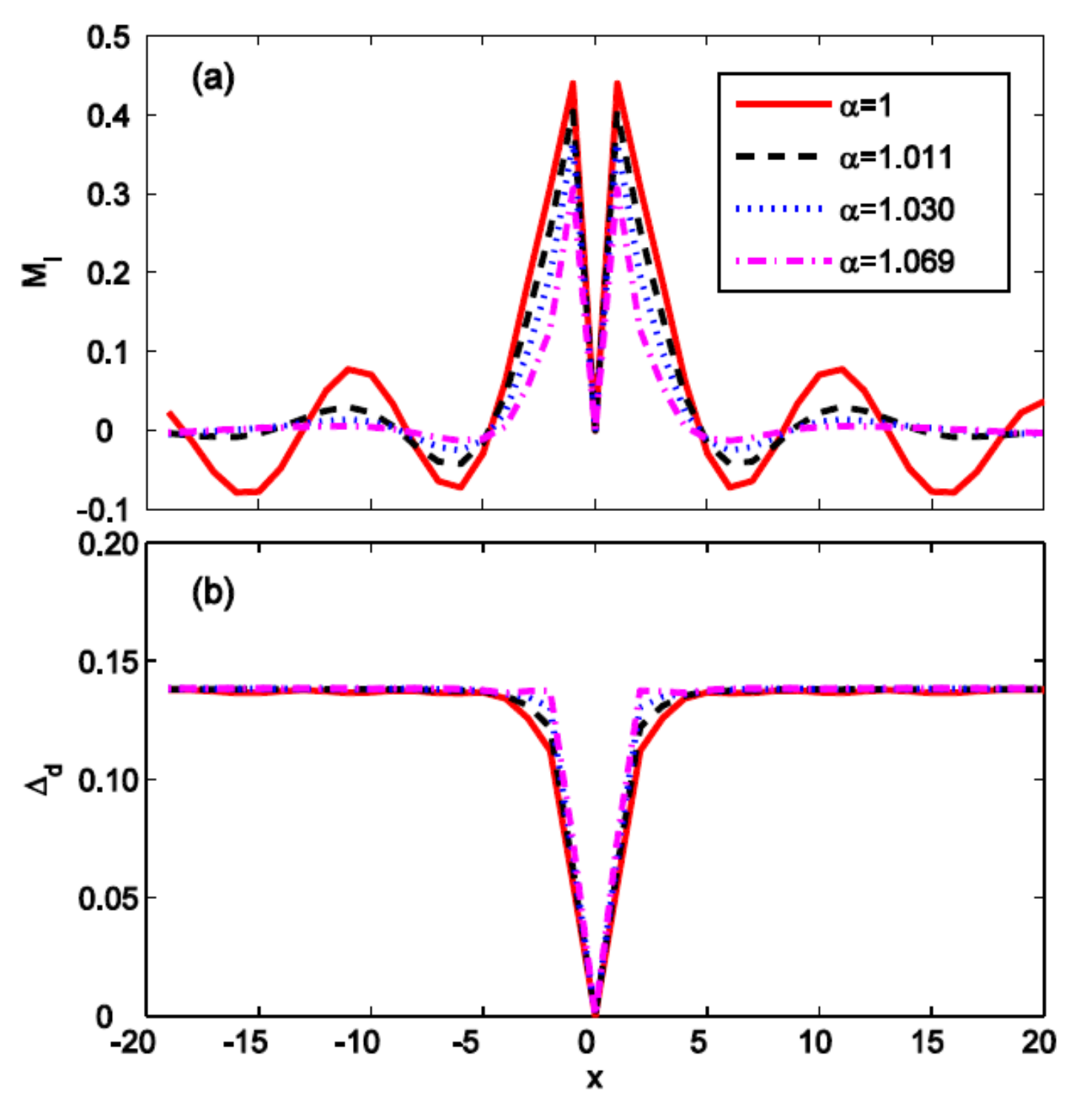}
\caption{Calculated spatial dependence of the competing magnetic $(M)$ and superconducting $(\Delta_d)$ order parameters around a unitary scattering centre in panels \textbf{a} and \textbf{b}, respectively. The abscissa is in units of the planar lattice parameter. The figures are based on an effective model with competing AFM and superconducting pairing interactions. The parameter  denotes a relative change in effective band width.}
\label{figure 7}
\end{figure} 
Figures 7a and b display the evolution of real-space texture of magnetic and superconducting orders around a unitary scattering centre for slightly different values of the effective bandwidth, which is in turn proportional to the external pressure. At ambient pressure ($\alpha=1$), the magnetic order is already induced around the impurity and exhibits an oscillatory behavior at quite a long range while the superconducting order is suppressed at the impurity site. With the increased bandwidth, the oscillating amplitude diminishes so that the magnetic order becomes more localized around the impurity, that is, a spin droplet is formed. Correspondingly, the superconducting order is recovered more quickly to the bulk value. Once the spin droplet is formed, its size also decreases with the bandwidth. This trend agrees qualitatively with the experimental measurements (see Fig.~6b). Though this mean-field model does not explicitly include quantum fluctuations in the Hamiltonian, it does include the ingredients to zeroth order of quantum spin fluctuations, from which higher-order ones can be calculated within the random phase approximation. Furthermore, the existence of spin droplets in Cd-doped CeCoIn$_5$ near the extrapolated QCP, as nicely captured in the present mean-field theory, is an important conclusion to be considered explicitly in a theory that goes beyond a mean-field approximation.

\end{document}